\documentclass[twocolumn,aps,pra,floatfix,showpacs,noshowkeys,epsfig,graphics]{revtex4}%
\usepackage{graphicx}
\usepackage{amsmath}
\usepackage{amsfonts}
\usepackage{amssymb}

\newcommand{\newc}{\newcommand}
\newc{\beq}    {\begin{equation}}
\newc{\eeq}    {\end{equation}}

\newc{\beqa}    {\begin{eqnarray}}
\newc{\eeqa}    {\end{eqnarray}}
\newc{\bs}    {\section}
\newc{\no}    {\\ \nonumber}

\begin{document}
\title{Quantum Separability of the vacuum for Scalar Fields with a Boundary}
\author{Jae-Weon Lee}\email{scikid@kias.re.kr}
\author{Jaewan Kim}\email{jaewan@kias.re.kr}
\affiliation{School of Computational Sciences,
             Korea Institute for Advanced Study,
             207-43 Cheongnyangni 2-dong, Dongdaemun-gu, Seoul 130-012, Korea}
\author{Taeseung Choi}\email{tchoi@korea.ac.kr}
\affiliation{Department of Physics,
             Korea university,Anam-dong Seongbuk-Gu, Seoul, 136-701, Korea}
\date{\today}
\begin{abstract}
Using  the Green's function approach we investigate separability  of the vacuum state of
a massless scalar field with a single Dirichlet boundary. Separability is demonstrated
using the positive partial transpose criterion for effective
two-mode Gaussian states of collective operators. In contrast to the vacuum energy,
entanglement of the vacuum is not modified by the presence of the
boundary.
\end{abstract}

%\keywords{entanglement; solid state qubits; quantum information}
\maketitle Recently theories and experiments in  quantum
information science\cite{Nielsen01,Vedral02} have been progressed
significantly. This has led to interest in studying  entanglement
in many-particle systems such as Bose Einstein
condensations\cite{bec}, Fermion systems\cite{Vedral03,oh},lattice systems\cite{narnhofer-2004-},
thermal boson systems\cite{braun-2002-89,my} and even in the
vacuum\cite{reznik,reznik-2003-33,reznik-2005-71}. Entanglement is now treated as a physical
quantity, as well as a resource for quantum information
processing. On the other hand, scalar  particles such as the Higgs
particles
 in the standard model are essential ingredients of modern physics.
 Although it is well known that the vacuums in some quantum field models have quantum nonlocality\cite{vacuumbell}, the existence of
entanglement of the general vacuum is still controversial.
 From the particle physics viewpoint, everything in the universe is made of quantum fields and
 massless non-interacting quantum scalar field is the simplest and,
 hence, the most basic quantum field. Furthermore, scalar fields can be  treated as an infinite-mode  extension of harmonic
  oscillators\cite{kofler:052107}.
  Therefore, studying
entanglement of the vacuum state of scalar fields  is very
important.  For decades the nature of the  vacuum, concerning the
Casimir energy\cite{casimir,casimir2}, in a bounded space has been
extensively studied using the Green's function
method\cite{dewitt}. In this paper, we use this   Green's function
approach  to investigate the  quantum entanglement of
the vacuum for non-interacting scalar fields  with an infinite Dirichlet boundary
(i.e., fields are constrained to vanish at the boundary). Since
the scalar fields are continuous variables we need a separability
criterion for continuous variables. Recently
there has been a renewal of interest in
entanglement of Gaussian states in the context of continuous
variable quantum information, and
entanglement measures such as the purity\cite{purity} or the
negativity has been suggested\cite{negativity} for two-mode
Gaussian states. In this work we shall average the fields over two
tiny boxes to make the vacuum state an effective two-mode Gaussian
state.

 We start by introducing the massless (real) Klein-Gordon  fields, which are described by a Lagrangian
 density, \beq
 \mathcal{L} (\vec{x},t)=\frac{1}{2} \partial_\alpha\phi
\partial^\alpha \phi,~~ ~(\alpha=t,x,y,z),
\eeq
which leads to the equation of motion $\partial_\alpha
\partial^\alpha \phi=0$
 in unbounded four-dimensional Minkowski spacetime.
In this  free-space (without a boundary) the quantum field operator of the scalar field can be expanded as
\beq
\phi(\vec{x},t)=\int \frac{d^3\vec{k}}{(2\pi)^3 \sqrt{2 \omega_{\vec{k}} }} \left(a(\vec{k})
 e^{i\vec{k}\cdot \vec{x}-i \omega_{k} t}+H.C. \right),
 \label{phi}
\eeq
where $\omega_k=|\vec{k}|$.
The equal time commutation relations of the scalar field are
\beqa
\label{commutation}
 [ \phi(\vec{x},t),\phi(\vec{x}',t)]&=&0 , [ \pi(\vec{x},t),\pi(\vec{x}',t)]=0, \no
 [ \phi(\vec{x},t), \pi(\vec{x}',t) ] &=& i \delta(\vec{x}-\vec{x}'),
 \eeqa
where $\pi(\vec{x},t)\equiv \partial_t \phi(\vec{x},t)$ is the
momentum operator for $\phi(\vec{x},t)$. Consider a vector
of the field and its momentum operator at two points $\vec{x}$ and
$\vec{x}'$
 at a time $t$, i. e., $\xi=(\phi(\vec{x},t),\pi(\vec{x},t),\phi(\vec{x}',t),\pi(\vec{x}',t))$.
The vacuum for the scalar field  ($|0\rangle$) has
 a
variance matrix having the following form:
 \beq
 V_{\alpha \beta}\equiv \frac{1}{2}\langle 0|\{\triangle \xi_\alpha,  \triangle \xi_\beta  \}|0\rangle,
 \eeq
 where $\{A,B\}=AB+BA$  and $\triangle \xi_\alpha\equiv \xi_\alpha-\langle 0| \xi_\alpha|0\rangle$
 with $\langle 0| \xi_\alpha|0\rangle=0$ in this paper.
 This matrix represents zero-temperature quantum fluctuation of the vacuum.
 Using the equal time commutation relations
 one can easily note that
$\langle 0| \xi_1\xi_3|0\rangle=\langle 0|\phi(\vec{x},t)\phi(\vec{x}',t)|0\rangle=\langle 0|\xi_3\xi_1|0\rangle$, hence
$V_{13}=V_{31}$.
For systems having  time-translational symmetry, like ours,
$\partial_t \langle 0|\phi(\vec{x},t)\phi(\vec{x},t)|0\rangle=0$.
Therefore, $\langle 0|\xi_1\xi_2|0\rangle=\langle 0|\phi(\vec{x},t)\pi(\vec{x},t)|0\rangle=-\langle 0|
\xi_2\xi_1|0\rangle$, and
$V_{12}=V_{21}=0$.
Hence the variance matrix has a following form
\begin{eqnarray}
V =            \begin{bmatrix}
            a & 0 & c  &0 \\
            0 & b & 0 &d \\
            c & 0 & a'  &0 \\
            0 & d & 0 & b'
           \end{bmatrix}\,
\end{eqnarray}
where $a=\langle0|\{\phi(\vec{x},t),\phi(\vec{x},t)\}|0\rangle/2$,
$b=\langle
0|\{\pi(\vec{x},t),\pi(\vec{x},t)\}|0\rangle/2, ~a'=\langle0|\{\phi(\vec{x'},t),\phi(\vec{x'},t)\}|0\rangle/2,$
and so on. Note that for the free-space field operator (Eq.
(\ref{phi})) $a$ and $a'$ term diverge.
The divergence, however, can be made disappear
if we consider a scalar field with a boundary. The
effect of the boundary can be calculated by  subtracting a
free-space Green's function from the Green's function with a
boundary (See  Eq.(\ref{GB}) and below). Divergent terms which are in the
both Green's functions cancel each other. The components of $V$
can be calculated from the Green's function called the Hadamard's
elementary function \cite{birrell}
 \beq
G(\vec{x},t;\vec{x}',t')=\langle 0|\{\phi(\vec{x},t),\phi(\vec{x}',t')\}|0\rangle.
 \eeq
Once we know the Green's function we can calculate the components of the variance matrix $V$ from it, i.e.,
 \beqa
 \label{cd}
c(\vec{x},\vec{x}')&=&\frac{1}{2} \langle
0|\{\phi(\vec{x},t),\phi(\vec{x}',t)\}|0\rangle \no
&=&\lim_{t\rightarrow
t'}\frac{1}{2} G(\vec{x},t;\vec{x}',t'),\no
d(\vec{x},\vec{x}')&=&\frac{1}{2}
\langle 0|\{\partial_t\phi(\vec{x},t),\partial_t\phi(\vec{x}',t)\}|0\rangle \no
&=&
\lim_{t\rightarrow t'} \partial_t \partial_{t'}
\frac{1}{2}G(\vec{x},t;\vec{x}',t').
 \eeqa
Then, $ a= \lim_{\vec{x}' \rightarrow \vec{x}}
c(\vec{x},\vec{x}')$, $ b= \lim_{\vec{x}'\rightarrow \vec{x}}$
$ d(\vec{x},\vec{x}')$ and so on.

With the boundary the mode expansion in Eq. (\ref{phi}) is no
longer valid. Instead one can directly work with a modified
Green's function to obtain $V$ as follows. First let us choose two
points $\vec{x}=(x,y,z)$ and $\vec{x'}=(x',y',z')$. Using the
method of images\cite{image} one can obtain a scalar field Green's
function $G_B$ with a Dirichlet boundary positioned at $z=0$
\cite{birrell};
 \beq
 G_B=G_0-\frac{1}{2\,{\pi }^2\,\left( -{\left(
t - {t'} \right) }^2 + r^2+ {\left( z + {z'} \right) }^2 \right) },
\label{GB}
\eeq
where  the free-space Green's function is $G_0=[ 2\,{\pi
}^2\,\left( -{\left( t - {t'}\right) }^2 + r^2 + {\left( z - {z'}
\right) }^2 \right) ]^{-1}$ and $r^2\equiv{\left( x - x' \right)
}^2 + {\left( y - y' \right) }^2$. By subtracting the free-space
Green's function from $G_B$ we obtain a regularized  Green's
function; $ G=G_B-G_0$ which is not divergent. From $G$ one can
obtain the components of $V$ using Eq. (\ref{cd});
 \beqa
  \label{component}
(a,b,a',b',c,d)= \frac{1}{2\pi^2}(
\frac{-1}{8\,{z^2}},\frac{1}{16\,z^4},
 \frac{-1}{8\,z'^2},\frac{1}{16\,\,z'^4}, \no
  \frac{-1}{2\,\left( r^2 +
       {\left( z + z' \right) }^2 \right) },
  \frac{1}{\,{\left( r^2+
         {\left( z + z' \right) }^2 \right) }^2}).
\eeqa
Now we discuss separability of the vacuum. Since `if and only
if ' separability test for infinite-mode states are unknown, we
need to
 reduce the infinite-mode states to effective two-mode Gaussian states.
 We follow
the approach in ref. \cite{kofler:052107}, i.e.,
  we spatially average the field operator over  two tiny boxes  centered at $\vec{x}$ and $\vec{x}'$, respectively
  (See Fig. 1).
  \begin{figure}[htbp]
\includegraphics[width=0.4\textwidth]{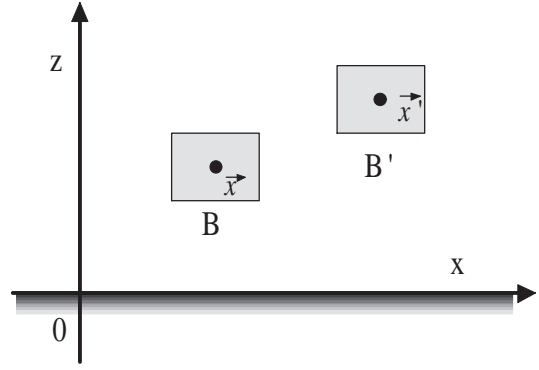}
\caption{
We average fields over two tiny boxes $B$ and $B'$  centered at $\vec{x}$ and $\vec{x}'$, respectively.
The Dirichlet boundary is at $z=0$.
 \label{ccorrelation} }
\end{figure}
Defining these collective operators is reasonable, because
in a physical situation probes always have  finite spatial resolution.
We, however, do not average but integrate the momentum operators within the box, since momentum is additive.
  Thus the collective operators are
  \beqa
\tilde{\xi}&\equiv&(\Phi(\vec{x},t),\Pi(\vec{x},t),\Phi(\vec{x}',t),\Pi(\vec{x}',t))\no
&=&(\frac{1}{L^3}\int_Bd^3\vec{y} \phi(\vec{x}+\vec{y},t),
  \int_Bd^3\vec{y} \pi(\vec{x}+\vec{y},t), \no
& & \frac{1}{L^3}\int_{B'}d^3\vec{y'} \phi(\vec{x}'+\vec{y'},t),
  \int_{B'}d^3\vec{y'} \pi(\vec{x}'+\vec{y'},t)).
  \eeqa

Here $\int_B d^3\vec{y} f(\vec{x}+\vec{y})$ denotes an integration of $f(\vec{x})$ over a box $B$ centered at $\vec{x}$ with volume $L^3$.
  Then, the commutation relations in Eq. (\ref{commutation})
  reduce to the  canonical commutation relations for the collective operators  \cite{kofler:052107}
\beq
[\tilde{\xi}_\alpha,\tilde{\xi}_\beta]=i\Omega_{\alpha\beta},
\eeq
where
\begin{eqnarray}
\Omega =            \begin{bmatrix}
            J & 0  \\
            0 & J \\
           \end{bmatrix}\,,
           J =            \begin{bmatrix}
            0 & 1  \\
            -1 & 0 \\
           \end{bmatrix}\,.
\end{eqnarray}
Similarly we define an variance matrix for the collective operators
$\tilde{V}_{\alpha \beta}\equiv \frac{1}{2}\langle 0|\{ \tilde{\xi}_\alpha,  \tilde{\xi}_\beta  \}|0\rangle.$

Now we show that, using the fundamental theorem of calculus\cite{calculus},
one can calculate the  variance matrix $\tilde{V}$ for the two-mode Gaussian states
with those of $V$ in the limit $L\rightarrow 0$ (but still non-zero).
For example,
\beqa
\tilde{V}_{13}&=&\lim_{L\rightarrow 0}\frac{1}{2}\langle 0|\{ \Phi(\vec{x},t),  \Phi(\vec{x'},t)\}|0\rangle \no
&=&\lim_{L\rightarrow 0} \frac{1}{2L^6}  \int_Bd^3\vec{y}\int_{B'}d^3\vec{y}'\langle 0|\{ \phi(\vec{x}+\vec{y},t),
 \phi(\vec{x'}+\vec{y}',t)\}|0\rangle \no
&=&\frac{1}{2}\langle 0|\{ \phi(\vec{x},t), \phi(\vec{x'},t)\}|0\rangle
=V_{13}=c.
\eeqa
This is possible since the integrand of the second line (i. e., $c$ term in Eq. (\ref{component}))
 is continuous on all the region $z>0$.  ( Due to the boundary, it is enough to consider only the
  half-space $z>0$.)
 Similarly
\beqa
\tilde{V}_{24}&=&
\lim_{L\rightarrow 0}  L^6\frac{1}{2L^6}  \int_Bd^3\vec{y}\int_{B'}d^3\vec{y}'\langle 0|\{ \pi(\vec{x}+\vec{y},t),\no
 &\pi(&\vec{x'}+\vec{y}',t)\}|0\rangle =L^6 V_{24}=L^6 d.
\eeqa
Hence,
\begin{eqnarray}
\tilde{V} =            \begin{bmatrix}
            a & 0 & c  &0 \\
            0 &L^6 b & 0 & L^6 d \\
            c & 0 & a'  &0 \\
            0 & L^6 d & 0 & L^6 b'
           \end{bmatrix}\,
\equiv
              \begin{bmatrix}
            A & G  \\
            G^T & B  \\
           \end{bmatrix}.
\end{eqnarray}
Obviously, our formalism does not work for $L=0$, but for $L$ infinitesimally small the above equation
gives an exact  result.
The separability  criterion we use in this paper is positive partial transpose (PPT) criterion\cite{peres,ppt}
for two-mode Gaussian states\cite{simon}
which is equivalent to
\beq
F\equiv \tilde{\Sigma}-(\frac{1}{4}+4 det \tilde{V})\leq 0,
\label{F}
\eeq
where $\tilde{\Sigma}=det A +det B-2det G$.
By inserting the components into Eq.(\ref{F}) we obtain
\beqa
F&=&-\frac{1}{4} \no &-&\frac{L^6}{2^2\pi^4}\left[\frac{1}{2^7 z^6}+\frac{1}{2^7 z'^6}-\frac{1}{[r^2+(z+z')^2]^3}\right] \no
&-&\frac{L^{12}}{2^6\pi^8}[\frac{1}{2^{12} z^6 z'^6}-\frac{1}{2^4 z^2 z'^2(r^2+(z+z')^2)^4}\no
&-&\frac{1}{2^8 z^4 z'^4 (r^2+(z+z')^2)^2}+
\frac{1}{ (r^2+(z+z')^2)^6}
]\no
&\le&-\frac{1}{4}.
\eeqa
One can obtain this inequality using two absolute inequalities,
$1/(X+Y)^n \le (1/X^n+1/Y^n)/2^{n+1}$ and $X^3+Y^3 \ge XY^2 +YX^2~(X,Y>0)$ on the two square brackets, respectively.
The maximum value is achieved when $\vec{x}=\vec{x} '~(r=0)$.
Therefore, the effective two-mode scalar field vacuum with a boundary is PPT and, hence, separable,
 when it is described by the variance matrix $\tilde{V}$ and the regularized Green's function $G$.
Our  results, however, do not  rule out the possibility  of
entanglement of the vacuum for  free-space scalar fields, since we
have subtracted the free-space Green's function which might give
entanglement. What our results really imply is that the presence
of a single boundary does not change the separability or the entanglement which the
vacuum for the free-space scalar fields may have,
 while the presence of the boundary modifies the vacuum
energy to $-1/16\pi^2 z^4$\cite{birrell}.

 Entanglement of the scalar field could be experimentally tested by the scheme with trapped ions\cite{reznik}
  or Bose-
Einstein condensates\cite{Kaszlikowski}.
 Our approach
provides a new method using the Hadamard's elementary function to
investigate entanglement of the quantum fields vacuum within a bounded space. It will be
interesting to investigate how the number of boundaries and
properties of the fields such as interactions, masses, charges, and
spins change the results.
\vskip 15pt
After completion of our work we
found that there appears a paper about spatial entanglement of
free thermal bosonic fields (quant-ph/0607069).
\vskip 15pt
\paragraph*{Acknowledgments.--}
\indent
We thank Prof.  C. Brukner, Prof. Jinhyoung Lee,  Prof. Jinho Cho, Prof. Mahn-Soo Choi and Dr. Hyunchul Nha
for helpful discussions.
J. Lee was supported by part by the Korea Ministry of Science and Technology.
J. Kim
was supported by the Korea Research Foundation (Grant
No. KRF-2002-070-C00029). T. Choi was supported by the SRC/ERC program of MOST/KOSEF
(R11-2000-071), the Korea Research Foundation Grant
(KRF-2005-070-C00055), the SK Fund, and the KIAS.

%\bibliography{entanglement}% Produces the bibliography via BibTeX.

\begin{thebibliography}{25}
\expandafter\ifx\csname natexlab\endcsname\relax\def\natexlab#1{#1}\fi
\expandafter\ifx\csname bibnamefont\endcsname\relax
  \def\bibnamefont#1{#1}\fi
\expandafter\ifx\csname bibfnamefont\endcsname\relax
  \def\bibfnamefont#1{#1}\fi
\expandafter\ifx\csname citenamefont\endcsname\relax
  \def\citenamefont#1{#1}\fi
\expandafter\ifx\csname url\endcsname\relax
  \def\url#1{\texttt{#1}}\fi
\expandafter\ifx\csname urlprefix\endcsname\relax\def\urlprefix{URL }\fi
\providecommand{\bibinfo}[2]{#2}
\providecommand{\eprint}[2][]{\url{#2}}

\bibitem[{\citenamefont{Nielsen and Chuang}(2001)}]{Nielsen01}
\bibinfo{author}{\bibfnamefont{M.~A.} \bibnamefont{Nielsen}} \bibnamefont{and}
  \bibinfo{author}{\bibfnamefont{I.~L.} \bibnamefont{Chuang}},
  \emph{\bibinfo{title}{Quantum Computation and Quantum Information}}
  (\bibinfo{publisher}{Cambridge University Press}, \bibinfo{year}{2001}).

\bibitem[{\citenamefont{Vedral}(2002)}]{Vedral02}
\bibinfo{author}{\bibfnamefont{V.}~\bibnamefont{Vedral}},
  \bibinfo{journal}{Rev. Mod. Phys.} \textbf{\bibinfo{volume}{74}},
  \bibinfo{pages}{197} (\bibinfo{year}{2002}).

\bibitem[{\citenamefont{Sorensen et~al.}(2001)\citenamefont{Sorensen, Duan,
  Cirac, and Zoller}}]{bec}
\bibinfo{author}{\bibfnamefont{A.}~\bibnamefont{Sorensen}},
  \bibinfo{author}{\bibfnamefont{L.-M.} \bibnamefont{Duan}},
  \bibinfo{author}{\bibfnamefont{J.~I.} \bibnamefont{Cirac}}, \bibnamefont{and}
  \bibinfo{author}{\bibfnamefont{P.}~\bibnamefont{Zoller}},
  \bibinfo{journal}{Nature} \textbf{\bibinfo{volume}{409}}, \bibinfo{pages}{63}
  (\bibinfo{year}{2001}).

\bibitem[{\citenamefont{Vedral}(2003)}]{Vedral03}
\bibinfo{author}{\bibfnamefont{V.}~\bibnamefont{Vedral}},
  \bibinfo{journal}{Central Eur. J. Phys.} \textbf{\bibinfo{volume}{1}},
  \bibinfo{pages}{289} (\bibinfo{year}{2003}).

\bibitem[{\citenamefont{Oh and Kim}(2004)}]{oh}
\bibinfo{author}{\bibfnamefont{S.}~\bibnamefont{Oh}} \bibnamefont{and}
  \bibinfo{author}{\bibfnamefont{J.}~\bibnamefont{Kim}},
  \bibinfo{journal}{Phys. Rev. A} \textbf{\bibinfo{volume}{69}},
  \bibinfo{pages}{054305} (\bibinfo{year}{2004}).

\bibitem[{\citenamefont{Narnhofer}(2004)}]{narnhofer-2004-}
\bibinfo{author}{\bibfnamefont{H.}~\bibnamefont{Narnhofer}}
  (\bibinfo{year}{2004}), \eprint{quant-ph/0412152}.

\bibitem[{\citenamefont{Lee et~al.}()\citenamefont{Lee, Oh, and Kim}}]{my}
\bibinfo{author}{\bibfnamefont{J.}~\bibnamefont{Lee}},
  \bibinfo{author}{\bibfnamefont{S.}~\bibnamefont{Oh}}, \bibnamefont{and}
  \bibinfo{author}{\bibfnamefont{J.}~\bibnamefont{Kim}},
  \eprint{quant-ph/0510137}.

\bibitem[{\citenamefont{Braun}(2002)}]{braun-2002-89}
\bibinfo{author}{\bibfnamefont{D.}~\bibnamefont{Braun}},
  \bibinfo{journal}{Phys. Rev. Lett.} \textbf{\bibinfo{volume}{89}},
  \bibinfo{pages}{277901} (\bibinfo{year}{2002}).

\bibitem[{\citenamefont{Retzker et~al.}(2005)\citenamefont{Retzker, Cirac, and
  Reznik}}]{reznik}
\bibinfo{author}{\bibfnamefont{A.}~\bibnamefont{Retzker}},
  \bibinfo{author}{\bibfnamefont{J.~I.} \bibnamefont{Cirac}}, \bibnamefont{and}
  \bibinfo{author}{\bibfnamefont{B.}~\bibnamefont{Reznik}},
  \bibinfo{journal}{Phys. Rev. Lett.} \textbf{\bibinfo{volume}{94}},
  \bibinfo{pages}{050504} (\bibinfo{year}{2005}).

\bibitem[{\citenamefont{Reznik}(2003)}]{reznik-2003-33}
\bibinfo{author}{\bibfnamefont{B.}~\bibnamefont{Reznik}},
  \bibinfo{journal}{Found.Phys.} \textbf{\bibinfo{volume}{33}},
  \bibinfo{pages}{167} (\bibinfo{year}{2003}).

\bibitem[{\citenamefont{Reznik et~al.}(2005)\citenamefont{Reznik, Retzker, and
  Silman}}]{reznik-2005-71}
\bibinfo{author}{\bibfnamefont{B.}~\bibnamefont{Reznik}},
  \bibinfo{author}{\bibfnamefont{A.}~\bibnamefont{Retzker}}, \bibnamefont{and}
  \bibinfo{author}{\bibfnamefont{J.}~\bibnamefont{Silman}},
  \bibinfo{journal}{Phys. Rev. A} \textbf{\bibinfo{volume}{71}},
  \bibinfo{pages}{042104} (\bibinfo{year}{2005}).

\bibitem[{\citenamefont{Summers and Werner}(1985)}]{vacuumbell}
\bibinfo{author}{\bibfnamefont{S.~J.} \bibnamefont{Summers}} \bibnamefont{and}
  \bibinfo{author}{\bibfnamefont{R.}~\bibnamefont{Werner}},
  \bibinfo{journal}{Phys. Lett. A} \textbf{\bibinfo{volume}{110}},
  \bibinfo{pages}{29} (\bibinfo{year}{1985}).

\bibitem[{\citenamefont{Kofler et~al.}(2006)\citenamefont{Kofler, Vedral, Kim,
  and Brukner}}]{kofler:052107}
\bibinfo{author}{\bibfnamefont{J.}~\bibnamefont{Kofler}},
  \bibinfo{author}{\bibfnamefont{V.}~\bibnamefont{Vedral}},
  \bibinfo{author}{\bibfnamefont{M.~S.} \bibnamefont{Kim}}, \bibnamefont{and}
  \bibinfo{author}{\bibfnamefont{C.}~\bibnamefont{Brukner}},
  \bibinfo{journal}{Phys. Rev. A} \textbf{\bibinfo{volume}{73}},
  \bibinfo{eid}{052107} (\bibinfo{year}{2006}).

\bibitem[{\citenamefont{Casimir}(1948)}]{casimir}
\bibinfo{author}{\bibfnamefont{H.}~\bibnamefont{Casimir}},
  \bibinfo{journal}{Proc. K. Ned. Akad. Wet.} \textbf{\bibinfo{volume}{51}},
  \bibinfo{pages}{793} (\bibinfo{year}{1948}).

\bibitem[{\citenamefont{Bordag et~al.}(2001)\citenamefont{Bordag, Mohideen, and
  Mostepanenko}}]{casimir2}
\bibinfo{author}{\bibfnamefont{M.}~\bibnamefont{Bordag}},
  \bibinfo{author}{\bibfnamefont{U.}~\bibnamefont{Mohideen}}, \bibnamefont{and}
  \bibinfo{author}{\bibfnamefont{V.}~\bibnamefont{Mostepanenko}},
  \bibinfo{journal}{Phys. Rept.} \textbf{\bibinfo{volume}{353}},
  \bibinfo{pages}{1} (\bibinfo{year}{2001}).

\bibitem[{\citenamefont{DeWitt}(1975)}]{dewitt}
\bibinfo{author}{\bibfnamefont{B.~S.} \bibnamefont{DeWitt}},
  \bibinfo{journal}{Phys. Rept.} \textbf{\bibinfo{volume}{19}},
  \bibinfo{pages}{295} (\bibinfo{year}{1975}).

\bibitem[{\citenamefont{Adesso et~al.}(2004)\citenamefont{Adesso, Serafini, and
  Illuminati}}]{purity}
\bibinfo{author}{\bibfnamefont{G.}~\bibnamefont{Adesso}},
  \bibinfo{author}{\bibfnamefont{A.}~\bibnamefont{Serafini}}, \bibnamefont{and}
  \bibinfo{author}{\bibfnamefont{F.}~\bibnamefont{Illuminati}},
  \bibinfo{journal}{Phys. Rev. Lett.} \textbf{\bibinfo{volume}{92}},
  \bibinfo{pages}{087901} (\bibinfo{year}{2004}).

\bibitem[{\citenamefont{Kim et~al.}(2002)\citenamefont{Kim, Lee, and
  Munro}}]{negativity}
\bibinfo{author}{\bibfnamefont{M.~S.} \bibnamefont{Kim}},
  \bibinfo{author}{\bibfnamefont{J.}~\bibnamefont{Lee}}, \bibnamefont{and}
  \bibinfo{author}{\bibfnamefont{W.~J.} \bibnamefont{Munro}},
  \bibinfo{journal}{Phys. Rev. A} \textbf{\bibinfo{volume}{66}},
  \bibinfo{pages}{030301} (\bibinfo{year}{2002}).

\bibitem[{\citenamefont{Birrell and Davies}(1982)}]{birrell}
\bibinfo{author}{\bibfnamefont{N.}~\bibnamefont{Birrell}} \bibnamefont{and}
  \bibinfo{author}{\bibfnamefont{P.}~\bibnamefont{Davies}},
  \emph{\bibinfo{title}{Quantum fields in curved space}}
  (\bibinfo{publisher}{Cambridge University Press}, \bibinfo{year}{1982}).

\bibitem[{\citenamefont{Brown and Maclay}(1969)}]{image}
\bibinfo{author}{\bibfnamefont{L.~S.} \bibnamefont{Brown}} \bibnamefont{and}
  \bibinfo{author}{\bibfnamefont{G.~J.} \bibnamefont{Maclay}},
  \bibinfo{journal}{Phys. Rev.} \textbf{\bibinfo{volume}{184}},
  \bibinfo{pages}{1272} (\bibinfo{year}{1969}).

\bibitem[{\citenamefont{Rudin}(1976)}]{calculus}
\bibinfo{author}{\bibfnamefont{W.}~\bibnamefont{Rudin}},
  \emph{\bibinfo{title}{Principles of Mathematical Analysis}}
  (\bibinfo{publisher}{McGraw-Hill}, \bibinfo{year}{1976}).

\bibitem[{\citenamefont{Peres}(1996)}]{peres}
\bibinfo{author}{\bibfnamefont{A.}~\bibnamefont{Peres}},
  \bibinfo{journal}{Phys. Rev. Lett.} \textbf{\bibinfo{volume}{77}},
  \bibinfo{pages}{1413} (\bibinfo{year}{1996}).

\bibitem[{\citenamefont{Horodecki et~al.}(1996)\citenamefont{Horodecki,
  Horodecki, and Horodecki}}]{ppt}
\bibinfo{author}{\bibfnamefont{M.}~\bibnamefont{Horodecki}},
  \bibinfo{author}{\bibfnamefont{P.}~\bibnamefont{Horodecki}},
  \bibnamefont{and}
  \bibinfo{author}{\bibfnamefont{R.}~\bibnamefont{Horodecki}},
  \bibinfo{journal}{Phys. Lett. A} \textbf{\bibinfo{volume}{223}},
  \bibinfo{pages}{1} (\bibinfo{year}{1996}).

\bibitem[{\citenamefont{Simon}(2000)}]{simon}
\bibinfo{author}{\bibfnamefont{R.}~\bibnamefont{Simon}},
  \bibinfo{journal}{Phys. Rev. Lett.} \textbf{\bibinfo{volume}{84}},
  \bibinfo{pages}{2726} (\bibinfo{year}{2000}).

\bibitem[{\citenamefont{Kaszlikowski and Vedral}()}]{Kaszlikowski}
\bibinfo{author}{\bibfnamefont{D.}~\bibnamefont{Kaszlikowski}}
  \bibnamefont{and} \bibinfo{author}{\bibfnamefont{V.}~\bibnamefont{Vedral}},
  \eprint{quant-ph/0606238}.

\end{thebibliography}

\end{document}